\documentclass[pdflatex,sn-mathphys-num]{sn-jnl}

\usepackage{graphicx}%
\usepackage{multirow}%
\usepackage{amsmath,amssymb,amsfonts}%
\usepackage{amsthm}%
\usepackage{mathrsfs}%
\usepackage[title]{appendix}%
\usepackage{xcolor}%
\usepackage{textcomp}%
\usepackage{manyfoot}%
\usepackage{booktabs}%
\usepackage{algorithm}%
\usepackage{algorithmicx}%
\usepackage{algpseudocode}%
\usepackage{listings}%
\usepackage{epsfig}
\usepackage{latexsym}
\usepackage{color}
\usepackage{xcolor}

\usepackage{xfrac}

\usepackage[english]{babel}
\usepackage{url}
\usepackage{needspace}
\usepackage{float}

\begin{document}

 \title{Systematics of the chemical freeze-out line in the high baryon density regime explored at SIS100}

\author[1,5]{\fnm{Emma Lilith} \sur{Hofmann}}

\author[7]{\fnm{Tom} \sur{Reichert}}%

\author[6]{\fnm{Volodymyr} \sur{Vovchenko}}

\author*[1,2]{\fnm{Jan} \sur{Steinheimer}}\email{j.steinheimer-froschauer@gsi.de}

\author[3,1,4]{\fnm{Marcus} \sur{Bleicher}}

\affil[1]{\orgname{GSI Helmholtzzentrum f\"ur Schwerionenforschung GmbH}, \orgaddress{\street{Planckstr. 1}, \city{Darmstadt}, \postcode{D-64291}, \country{Germany}}}

\affil[2]{\orgname{Frankfurt Institute for Advanced Studies}, \orgaddress{\street{Ruth-Moufang-Str. 1}, \city{Frankfurt am Main}, \postcode{60438}, \country{Germany}}}

\affil[3]{\orgdiv{Institut für Theoretische Physik}, \orgname{Goethe-Universit\"{a}t Frankfurt}, \orgaddress{\street{Max-von-Laue-Str. 1}, \city{Frankfurt am Main}, \postcode{60438}, \country{Germany}}}

\affil[4]{\orgname{Helmholtz Research Academy Hesse for FAIR (HFHF), GSI Helmholtzzentrum f\"ur Schwerionenforschung GmbH}, \orgaddress{\street{Max-von-Laue-Str. 12}, \city{Frankfurt am Main}, \postcode{60438}, \country{Germany}}}

\affil[5]{\orgname{Liebfrauenschule Bensheim}, \orgaddress{\street{ Obergasse 38}, \city{Bensheim}, \postcode{64625}, \country{Germany}}}

\affil[6]{\orgname{Department of Physics, University of Houston}, \orgaddress{\street{3507 Cullen Blvd}, \city{Houston}, \postcode{77204}, \country{USA}}}

\affil[7]{\orgdiv{Department of Physics}, \orgname{Duke University}, \orgaddress{\city{Durham}, \postcode{NC 27708}, \country{USA}}}


\abstract{
The systematic uncertainties of chemical freeze-out fits at SIS100 energies (Au+Au reactions at $\sqrt{s_{NN}}=3-5$ GeV) are studied using UrQMD simulations. Although hadron production in UrQMD does not occur on a sharp chemical freeze-out hyper-surface, the extracted fit quality is shown to be very good. The extracted chemical parameters depend on the selected hadron species as well as the underlying equation of state (EoS) of the matter. Including light nuclei and anti-protons in the fit increases the expected freeze-out temperature, while a stiffer EoS increases the obtained chemical potential. Similarly, the baryon densities extracted by the thermal fits depend on the choice of hadrons as well as the underlying equation of state. These results are important for the upcoming CBM@FAIR physics program and highlight that a degree of caution is advised when one relates the chemical freeze-out curve to features on the QCD phase diagram like the critical endpoint or a possible phase transition.}

\maketitle

\section{Introduction}
Collisions of atomic nuclei at high energies offer unique access to strongly interacting matter. The underlying theory, Quantum Chromodynamics (QCD), can therefore be studied under extreme conditions in a controlled laboratory setting \cite{Sorensen:2023zkk}. A very prominent feature of QCD is its asymptotic freedom \cite{Gross:1973id} which has lead to the hypothesis that a plasma-like state of quasi-free quarks and gluons is created at large temperatures and/or chemical potentials \cite{Cabibbo:1975ig,Rischke:2003mt}. Ab initio calculations of a discretized version of QCD on a lattice (lQCD) provide access to the phase diagram and have revealed that at zero chemical potential the transition from hadronic matter to the Quark-Gluon Plasma (QGP) is a crossover transition \cite{Borsanyi:2010bp,Bazavov:2011nk,Bazavov:2017dus}. It is a widely accepted, but still debated, belief that the phase transition has a critical point where the order of the phase transition changes from first to second order. However, the location (and existence) of this critical end point (CEP) remains unknown.

Extrapolations of lattice QCD calculations have ruled out the existence of a CEP up to $\mu_\mathrm{B}/T=3$ \cite{Borsanyi:2020fev,HotQCD:2018pds,Vovchenko:2017gkg}. Recently, many theoretical efforts based on different modeling choices and approaches seem to converge to a region for the CEP of $T=80-120$ MeV and $\mu_\mathrm{B} = 500-600$ MeV \cite{Fischer:2014ata,Fu:2019hdw,Gao:2020fbl,Gunkel:2021oya,Hippert:2023bel,Basar:2023nkp,Clarke:2024seq,Sorensen:2024mry,Shah:2024img,Ecker:2025vnb}, however, not all approaches agree on this region (see e.g. \cite{Eser:2023oii,Steinheimer:2025hsr}) or even the CEP's existence.

In a simplified picture for the evolution of a heavy-ion collision, the (almost) equilibrium expansion stage ends with a so-called chemical freeze-out at which flavor-changing reactions are thought to cease and the chemical composition that can be observed by experiments is fixed. After that, a kinetic freeze-out, at which all remaining reactions cease and particle spectra are fixed, occurs. Assuming chemical equilibrium up to the stage of chemical freeze-out, allows to use (grand-)canonical thermal models to extract the temperature and chemical potentials, at that final emission stage, from measured hadron multiplicities. A similar feature is also present in transport simulations \cite{Bleicher:2002dm,Reichert:2020yhx} and is understood as being due to the interplay of energy dependent (microscopic) scattering rates and the system's expansion. Thermal model fits to measured experimental data over a variety of collision systems and over 3 orders of magnitude in beam energy \cite{Cleymans:1998fq,Becattini:2000jw,Cleymans:2005xv,Andronic:2005yp,HADES:2010wua,Lorenz:2014eja,Becattini:2016xct,Lysenko:2024hqp} have lead to the chemical freeze-out line. Its name comes from the fact that almost all the chemical freeze-out points found in these fits lie along a very narrow band in the chemical potential - temperature plane. 
Typically these model fits use $\chi^2/\mathrm{d.o.f.}$ to quantify the fit quality. However, as has e.g. been shown in \cite{Motornenko:2021nds} considering different sets of hadrons during fitting may lead to degenerate minima, hence multiple equally good descriptions of the measured data. There are also calculations pointing towards differences in strangeness decoupling \cite{Bellwied:2013cta,Alba:2014eba,Bluhm:2018aei,Stafford:2019yuy,Flor:2020fdw,Reichert:2022qvt}, i.e. a flavor hierarchy.

This chemical freeze-out line is important as it may provide a reference for the location of important features in the phase diagram of QCD like a critical endpoint and associated phase transition. In addition, only the assumption that some level of local equilibrium can be achieved in high energy collisions would allow the interpretation of these collisions in terms of macroscopic quantities and an equation of state.

As an example, the proximity of the thermal model fits at vanishing chemical potential to the crossover region of QCD, established by lattice QCD calculations, hints to a chemical decoupling directly from the QGP, but such interpretation is still debated \cite{Reichert:2020yhx}. Furthermore, any arguments about observables that rely on the freeze-out line being close to a potential critical endpoint need to take into account the systematic uncertainties that are involved when determining this line \footnote{In addition to the fact that a heavy ion collisions should not be considered a homogeneous system of one fixed temperature at any given time but an ensemble of highly fluctuating densities.}. 

In this article we will discuss the systematics of determining the freeze-out line in the SIS100 and RHIC BESII beam energy range which has become the center of attention in the search for the critical endpoint and the QCD phase transition.
To do so, we calculate final state hadron multiplicities with the state-of-the-art Ultra-relativistic Quantum Molecular Dynamics (UrQMD) transport model. We will then use Thermal-FIST to fit the calculated hadron yields obtained from the UrQMD simulations using different hadron sets, thus following the common procedure to extract $T-\mu_\mathrm{B}$ points in the phase diagram. The goal is to determine the systematic dependencies of the extracted freeze-out line to the chosen set of hadrons as well as the underlying equation of state of the model which will give us insights into the uncertainties of the freeze-out line.

\section{UrQMD and Thermal-FIST}
\subsection{Ultra-relativistic Quantum Molecular Dynamics (UrQMD)}
The Ultra-relativistic Quantum Molecular Dynamics (UrQMD v4.0) model \cite{Bass:1998ca,Bleicher:1999xi,Bleicher:2022kcu} is a relativistic transport model with hadronic degrees of freedom covering roughly 100 hadronic species and their resonances. Binary interactions are included via the geometrical interpretation of the cross section. The cross sections are taken from experimental data wherever available or derived from effective models. At higher energies string excitation and their subsequent fragmentation are incorporated. In its QMD mode, the particles' trajectories between scattering are calculated solving the non-relativistic equations of motion including density- and momentum-dependent potentials. The momentum dependent potentials used are based on the interactions in a parity doublet chiral mean field model \cite{Papazoglou:1998vr,Steinheimer:2010ib,Motornenko:2019arp} and the parameters are the same as in \cite{Steinheimer:2024eha,Steinheimer:2025hsr} where a detailed description of the model can be found. 
In the present study we have employed both the cascade version of UrQMD (i.e. without potential interaction) and the QMD mode. The cascade mode provides an equation of state that is much softer than the QMD mode with the CMF potentials. As shown in a previous study \cite{Steinheimer:2025hsr} a different equation of state will lead to modifications of the hadron yields in particular of strange hadrons \cite{Hartnack:1993bq,Hartnack:1993bp,Fuchs:2005zg,Hartnack:2005tr,Hartnack:2011cn}. 

\begin{figure} [t]
    \centering
    \includegraphics[width=0.6\textwidth]{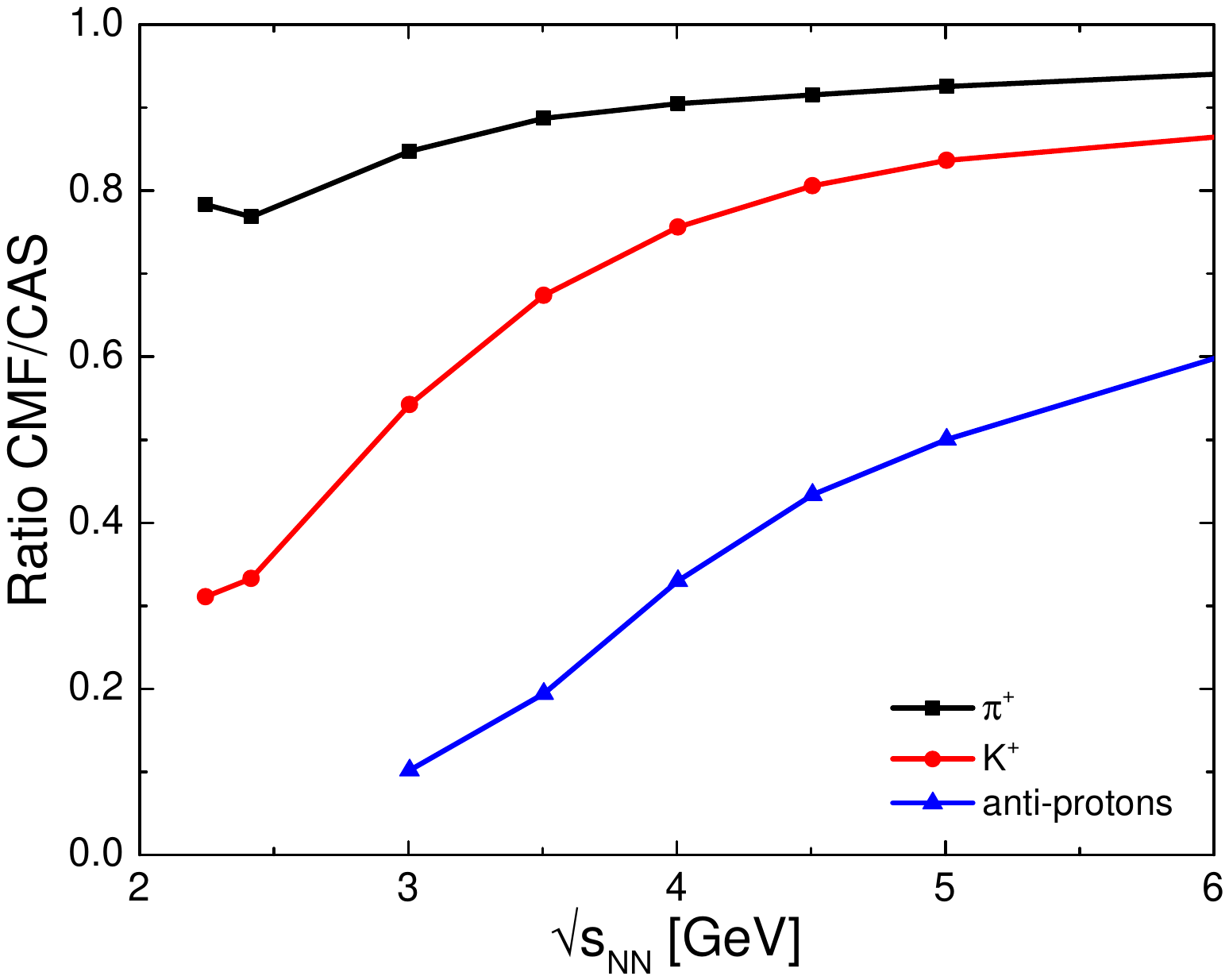}
    \caption{Ratios of full acceptance hadron yields of $\pi^+$ (black squares, solid lines), $K^+$ (red circles, solid lines) and $\Bar{p}$ (blue triangles, solid lines) as a function of $\sqrt{s_{NN}}$ from UrQMD simulations with CMF potentials with respect to cascade mode (CAS).}
    \label{fig:ratios}
\end{figure}

Figure \ref{fig:ratios} shows the ratios of integrated (i.e. in full acceptance) hadron multiplicities of UrQMD simulations for central ($0-10\%$) Au+Au collisions from $\sqrt{s_{NN}}=2-6$ GeV corresponding to the energy range of SIS100. We take the ratios of calculations with the CMF-potentials over the results using the cascade mode (CAS) of UrQMD. 
Shown are selected hadron yield ratios of $\pi^+$ (black squares, solid lines), $K^+$ (red circles, solid lines) and $\Bar{p}$ (blue triangles, solid lines).
As one can see especially strange hadrons, but also anti-baryons are significantly suppressed in the scenario with a stiffer EoS (i.e. with CMF potentials included) due to the reduced compression achieved. As we will see later these differences will be reflected in the chemical freeze-out parameters. In addition to stable hadrons the UrQMD results also include the multiplicities of deuterons which are calculated via a well tested coalescence algorithm \cite{Sombun:2018yqh,Bumnedpan:2024jit}. Note, that the coalescence procedure includes light nuclei up to $^4$He as well as hypernuclei and conserves baryon number as well as charge and strangeness.

\subsection{Thermal, Fast and Interactive Statistical Toolkit (Thermal-FIST)}
We employ the Thermal-FIST framework (Thermal, Fast and Interactive Statistical Toolkit) \cite{Vovchenko:2019pjl}. It allows to calculate the thermodynamics of hadronic matter within the hadron resonance gas (HRG) approximation. It was shown that the HRG model successfully describes the bulk properties of strongly interacting matter in the hadronic phase and allows to extract chemical properties (temperature and chemical potentials) to interpret hadron yield data from heavy-ion collisions \cite{Andronic:2017pug, Becattini:2016xct}. Thermal-FIST incorporates several extensions of the ideal HRG model (based usually on a grand canonical ensemble), including excluded-volume corrections, van der Waals-type interactions between baryons, finite resonance widths, and canonical suppression effects. It supports calculations in the grand-canonical, canonical, and mixed-canonical ensembles, enabling detailed studies of conserved-charge fluctuations and correlations relevant for exploring the QCD phase diagram \cite{Vovchenko:2015pya, Vovchenko:2017drx}. 

For the present investigation the grand canonical ensemble (GCE), including light nuclei in the particle list, is used to fit the data, while strangeness suppression is modeled using a standard $\gamma_s$ suppression parameter. The conditions of (average) strangeness neutrality and a total charge to baryon ration of $Q/A=0.4$ is enforced by introducing the strange and electric-charge chemical potentials $\mu_S$ and $\mu_Q$ as fit parameters. \footnote{ To check, we have also applied the strangeness canonical version of the model and found that it provides very similar results.} To perform $\chi^2$ fits one also needs to provide (statistical) error bars. While the statistics in the simulations can in principle be increased arbitrarily, minimizing the statistical error, here we nevertheless assume a $10\%$ statistical error for all hadronic species, similar to the typical systematic error values found in the experiment. This avoids introducing a bias to extremely rare particle species. For all the fits in the following we always followed the 'standard' procedure and employed the HRG model without any modifications to the EoS.

\section{Results}
Let us begin by a study of the 'missing hadrons' effect. This means we explore how the fit results change, if certain types of hadrons are not included in the fitting procedure. Reasons for this can be either (most of the time unwanted) experimental restrictions, like limited statistics or difficulties in the measurement (e.g. neutral final state particles) or to explore certain physics ideas, like the speculated flavor freeze-out hierarchy. Three different sets of hadrons have been selected for each equation of state:
\begin{itemize}
    \item Set 1: $\pi^+$, $\pi^-$, $K^+$, $K^-$, $p$, $\Lambda$, $\Bar{p}$,  and deuterons
    \item Set 2: $\pi^+$, $\pi^-$, $K^+$, $K^-$, $p$, $\Lambda$, $\Bar{p}$ 
    \item Set 3: $\pi^+$, $\pi^-$, $K^+$, $K^-$, $p$, $\Lambda$
\end{itemize}
where we will refer to set 1 as ``all hadrons''.

\begin{figure} [t]
    \centering
    \includegraphics[width=0.6\textwidth]{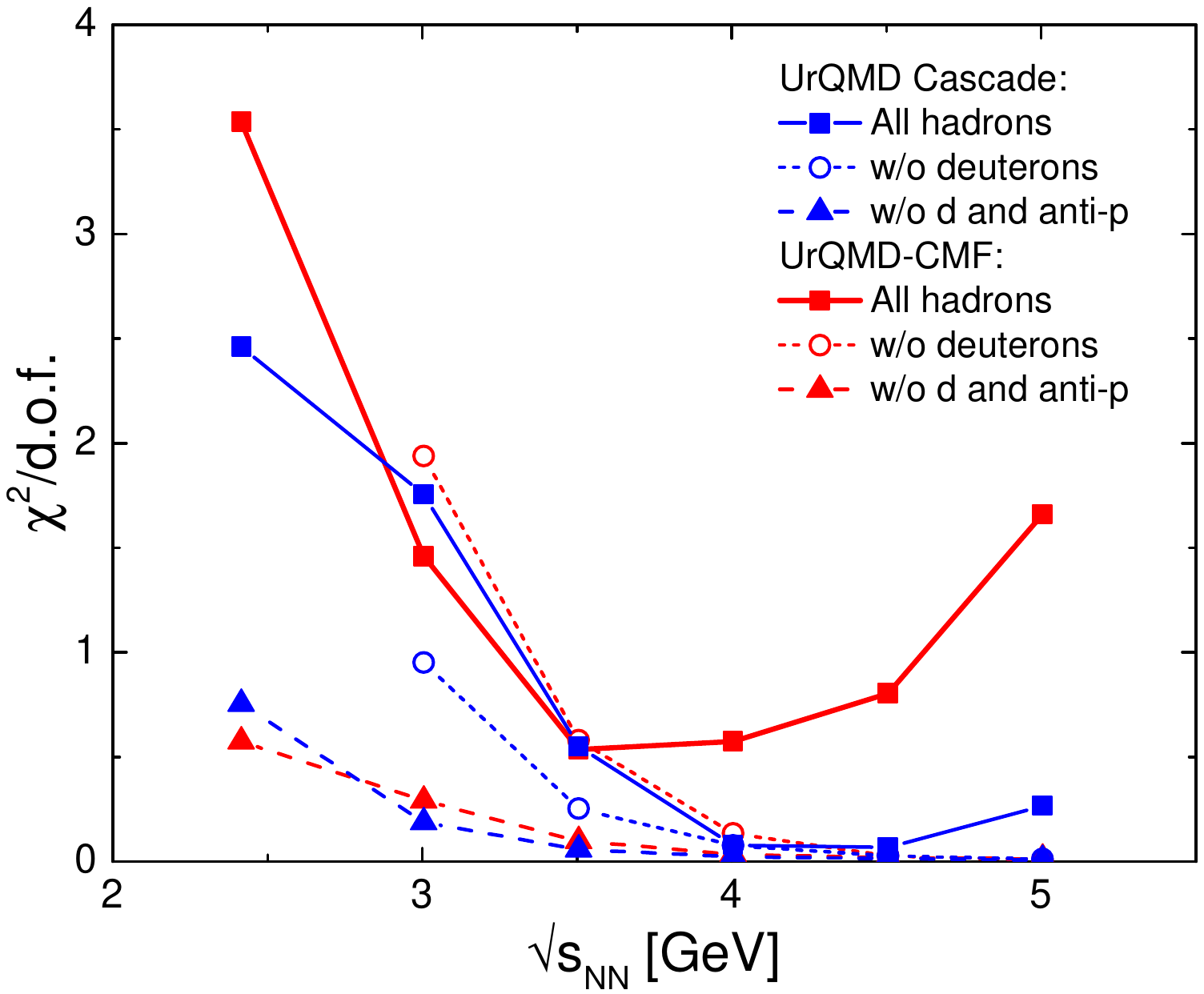}
    \caption{Fit quality given by $\chi^2/\mathrm{d.o.f.}$ for the different scenarios as a function of $\sqrt{s_{NN}}$. The blue curves show UrQMD calculations in cascade mode (CAS), red curves show UrQMD with CMF potentials. The full squares with solid lines show the fit quality including all hadrons (Set 1), open circles with dotted lines exclude the deuteron (Set 2) and full triangles with dashed lines exclude both, the deuteron and the anti-proton (Set 3).}
    \label{fig:chi2}
\end{figure}

Thermal-FIST is then used to fit hadron sets 1-3 for both scenarios of the UrQMD calculations and the resulting fit quality is investigated. Fig. \ref{fig:chi2} shows the fit quality quantified by $\chi^2/\mathrm{d.o.f.}$ for the different scenarios as a function of $\sqrt{s_{NN}}$. The blue curves show UrQMD calculations in cascade mode (CAS), red curves show UrQMD with CMF potentials. In both cases the full squares with solid lines show the fit quality including all hadrons (Set 1), open circles with dotted lines exclude the deuteron (Set 2) and full triangles with dashed lines exclude both, the deuteron and the anti-proton (Set 3). In general the fit quality is very good, with a $\chi^2$ per d.o.f. mostly below 1. Only in the scenario with deuterons (Set 1), the fit quality is not optimal for lower and higher beam energies. This can be understood as a result of the simplified coalescence description which becomes worse for the lower beam energies in full acceptance and the fact that the model does not well describe baryon stopping at higher beam energies which translates into an insufficient description of deuterons. It is important to note that even though the simulations with UrQMD themselves do not assume any form of equilibrium, the resulting hadron multiplicities for both UrQMD scenarios can be well approximated with an equilibrium fit. Thus, a good quality fit does not necessarily imply equilibrium.

Even though the underlying equation of state used in the simulation with the CMF equation of state (leading to changes in the particle yields) is not consistent with the equation of state used for the fitting, a satisfactory fit can still be obtained. Similarly, a HRG fit to real data may give a good fit result but not the 'correct' T-$\mu_B$-values.

\begin{figure*} [t]
    \centering
    \includegraphics[width=0.49\textwidth]{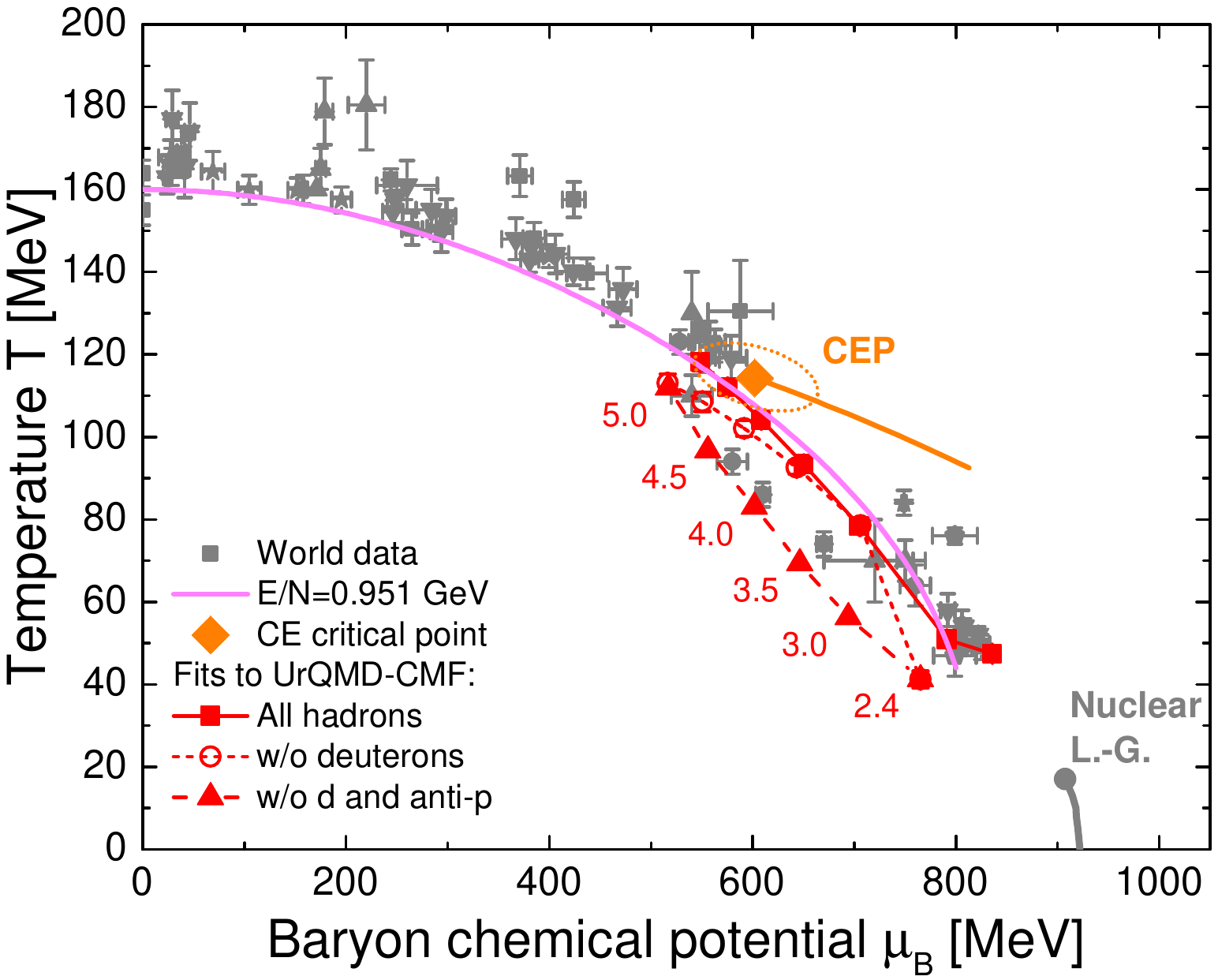}
    \includegraphics[width=0.49\textwidth]{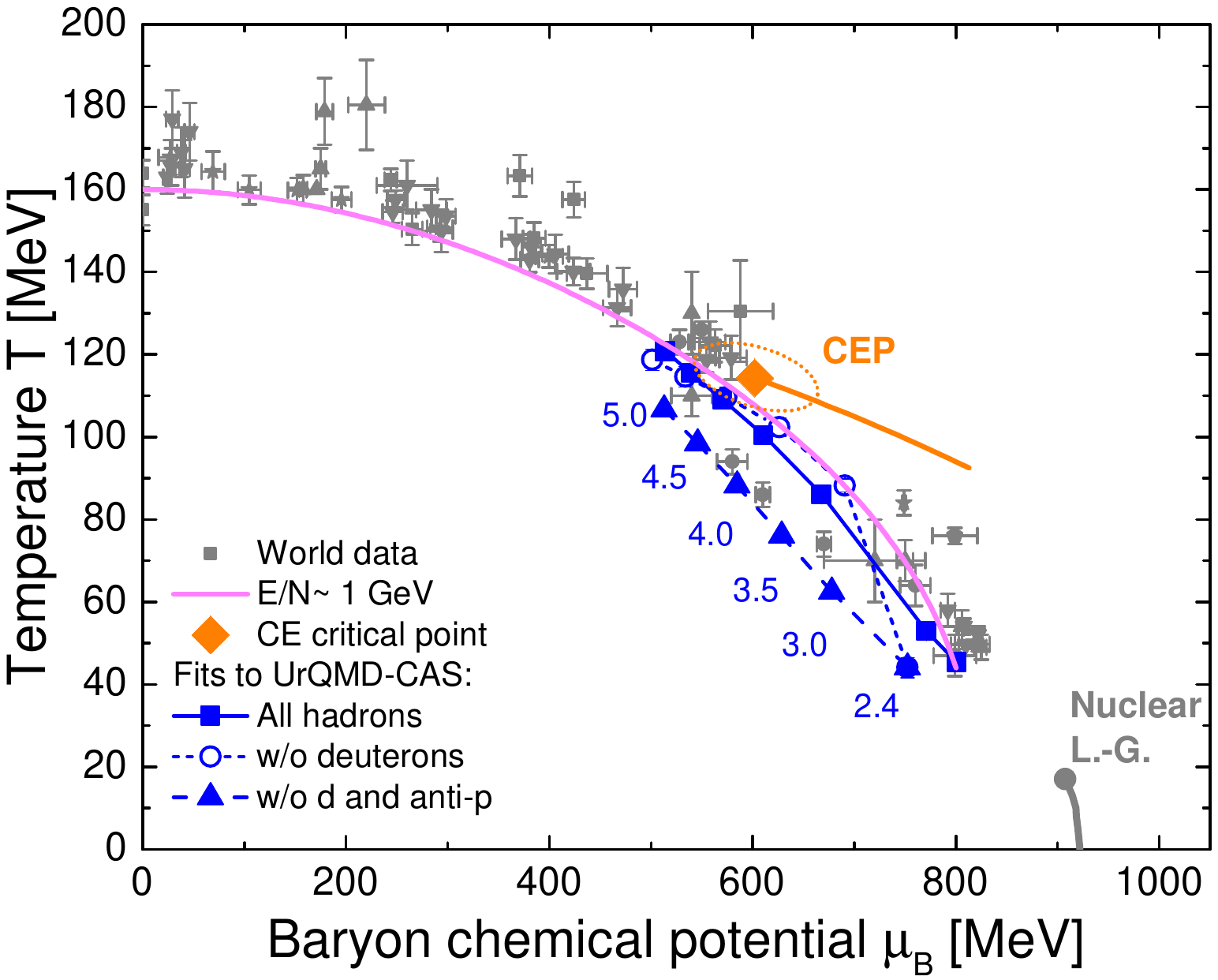}
    \caption{ Extracted chemical freeze-out curves in the $T-\mu_\mathrm{B}$ plane from UrQMD simulations with CMF potentials (left panel) and UrQMD in cascade mode (right panel). Three scenarios are compared: all hadrons (Set 1, full squares, solid lines), excluding deuterons from the fit (Set 2, open circles, dotted lines) and also excluding anti-protons (Set 3, full triangles, dashed lines). T,$\mu_B$ values obtained from fits to experimental data are shown as grey symbols \cite{Cleymans:1998fq,Becattini:2000jw,Cleymans:2005xv,Andronic:2005yp,HADES:2010wua,Lorenz:2014eja,Becattini:2016xct,Lysenko:2024hqp}. The magenta lines correspond to a constant energy per hadron line as a proxy for chemical freeze-out line, both for vanishing net-strangenss and $Q/A=0.4$ conditions. 
    The CEP shown as orange symbol with transition line is taken from \cite{Shah:2024img} and does not assume strangeness neutrality.}
    \label{fig:curves_mu}
\end{figure*}

Fig. \ref{fig:curves_mu} shows the extracted chemical freeze-out curves in the $T-\mu_\mathrm{B}$ plane from UrQMD simulations with CMF potentials (left panel) and UrQMD in cascade mode (CAS, right panel). Again, we compare the three sets of hadrons: all hadrons (Set 1, full squares, solid lines), excluding deuterons from the fit (Set 2, open circles, dotted lines) and also excluding anti-protons (Set 3, full triangles, dashed lines). The corresponding center of mass energies are also indicated next to the symbols.

The magenta line shows the parametrized freeze-out curve from \cite{Lysenko:2024hqp} for strangeness neutrality and a fixed $Q/A=0.4$. The extracted UrQMD freeze-out points follow this line closely. The existing world data of thermal fits to various different experiments is shown as grey symbols. These show that especially in the SIS100 beam energy range still some tension between existing data and the parametrized line exists.

To emphasize the relevance of this beam energy range for probing the QCD phase structure, we also show the prediction of a possible critical endpoint location from an expansion of constant entropy contours, using lattice QCD results as input \cite{Shah:2024img}, as an orange symbol with the associated transition line.\footnote{Note that this point and line was calculated assuming $\mu_S=0$ and $\mu_Q=0$ so a direct comparison should be done with care.} 
Similar predictions in recent literature also come from other approaches, such as functional methods~\cite{Gao:2020fbl,Gunkel:2021oya} or holographic models~\cite{Hippert:2023bel,Ecker:2025vnb}.

We can observe a clear difference between the fit parameters from the three hadron sets. In general the fit temperature increases, when deuterons and anti-protons are included in the fit (i.e. from Set 3 to Set 1). This can be understood, as the anti-protons are produced very early. The trend towards higher temperatures from including light nuclei has also been observed earlier. 

In addition, we also observe a small shift of the freeze-out points along the freeze-out line when a different equation of state is used. This corresponds to a shift in the entropy per baryon that is created in the collision. The simulations with the stiffer CMF EoS tend to show a smaller entropy per baryon than the cascade simulations. 

\begin{figure*} [t]
    \centering
    \includegraphics[width=0.49\textwidth]{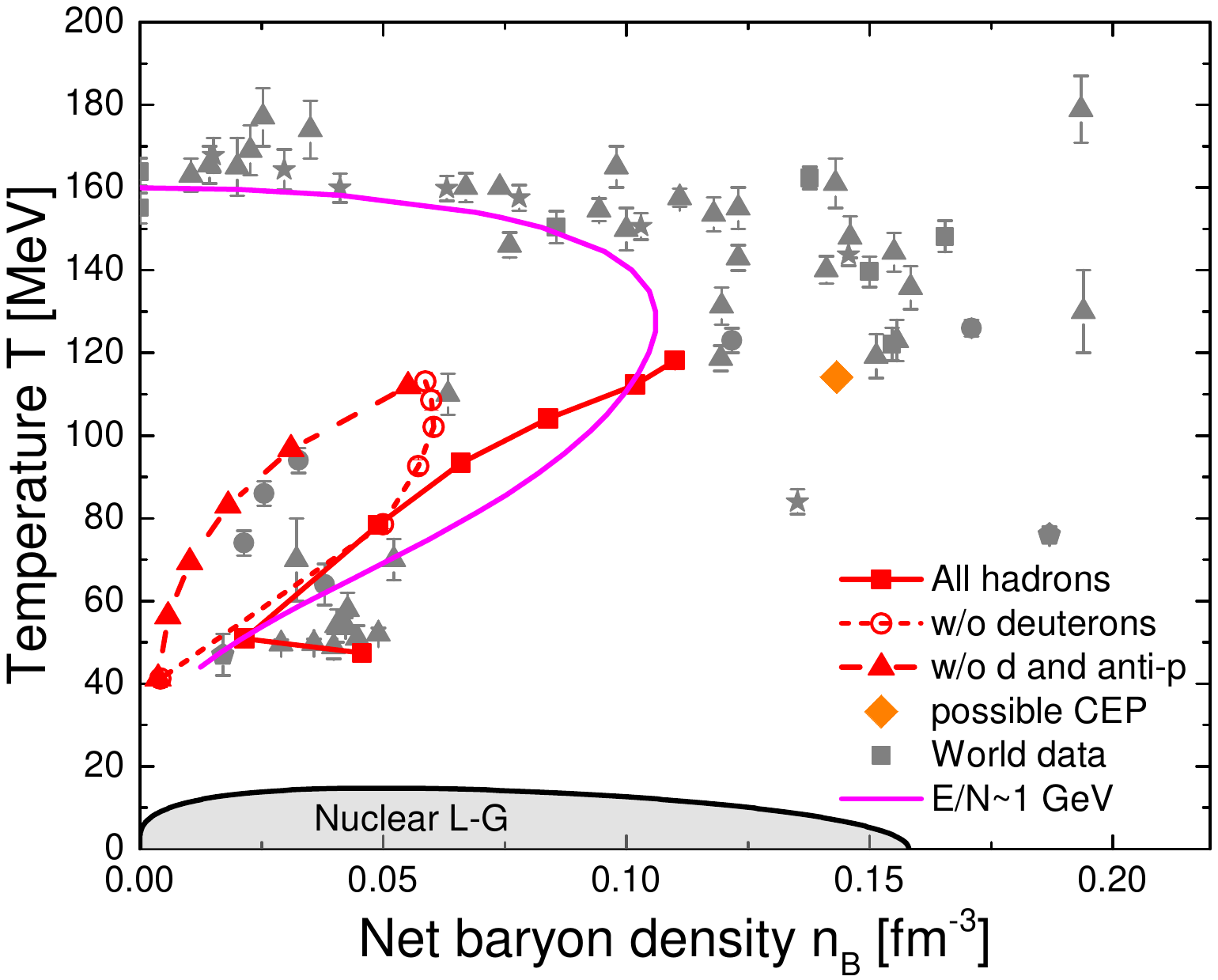}
    \includegraphics[width=0.49\textwidth]{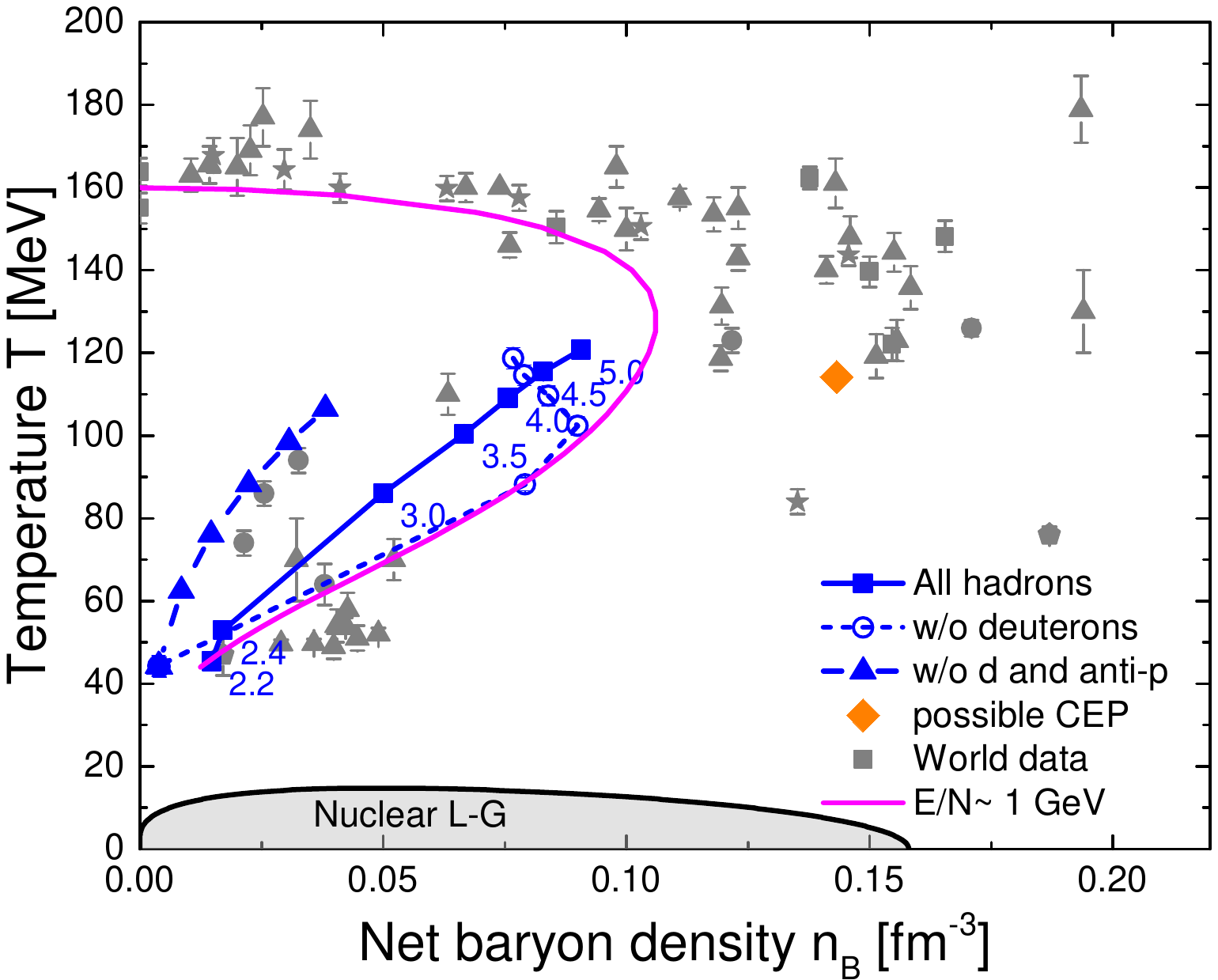}
    \caption{Extracted chemical freeze-out curves in the $T-n_\mathrm{B}$ plane from UrQMD simulations with CMF potentials (left) and UrQMD in cascade mode (right). Three scenarios are compared: all hadrons (full squares, solid lines), excluding deuterons from the fit (open circles, dotted lines) and also excluding anti-protons (full triangles, dashed lines). Densities have been computed assuming an ideal HRG and the thermal parameters from the fits enforcing vanishing net-strangeness. The orange diamond symbol depicts the location of the CEP shown in figures \ref{fig:curves_mu} for comparison, where the net baryon density was calculated assuming $\mu_S=0$ and $\mu_Q=0$.
    }
    \label{fig:curves_rho}
\end{figure*}

Fig. \ref{fig:curves_rho} shows the extracted chemical freeze-out curves in the $T-n_\mathrm{B}$ plane from UrQMD simulations with CMF potentials (left) and UrQMD in cascade mode (right). Again, three scenarios are compared: all hadrons (full squares, solid lines), excluding deuterons from the fit (open circles, dotted lines) and also excluding anti-protons (full triangles, dashed lines). Densities have been computed the thermal FIST which assumes an ideal HRG equation of state and using the thermal parameters from fit. The magenta line shows again the parametrized freeze-out curve results from \cite{Lysenko:2024hqp}.

In this representation of the phase diagram, the freeze-out curves differ more. An important result is that for the SIS100 energy regime both temperature and net baryon density at the freeze-out point appear to be increasing with beam energy. At some point this increase will saturate and the net baryon density will decrease again due to the reduced baryon stopping. However, this point is not yet reached even for the highest SIS100 beam energy. As before, we also observe a difference for the two EoS scenarios. The stiffer CMF-EoS leads to larger freeze-out densities and higher temperatures.

\begin{figure*} [t]
\centering
    \includegraphics[width=0.49\textwidth]{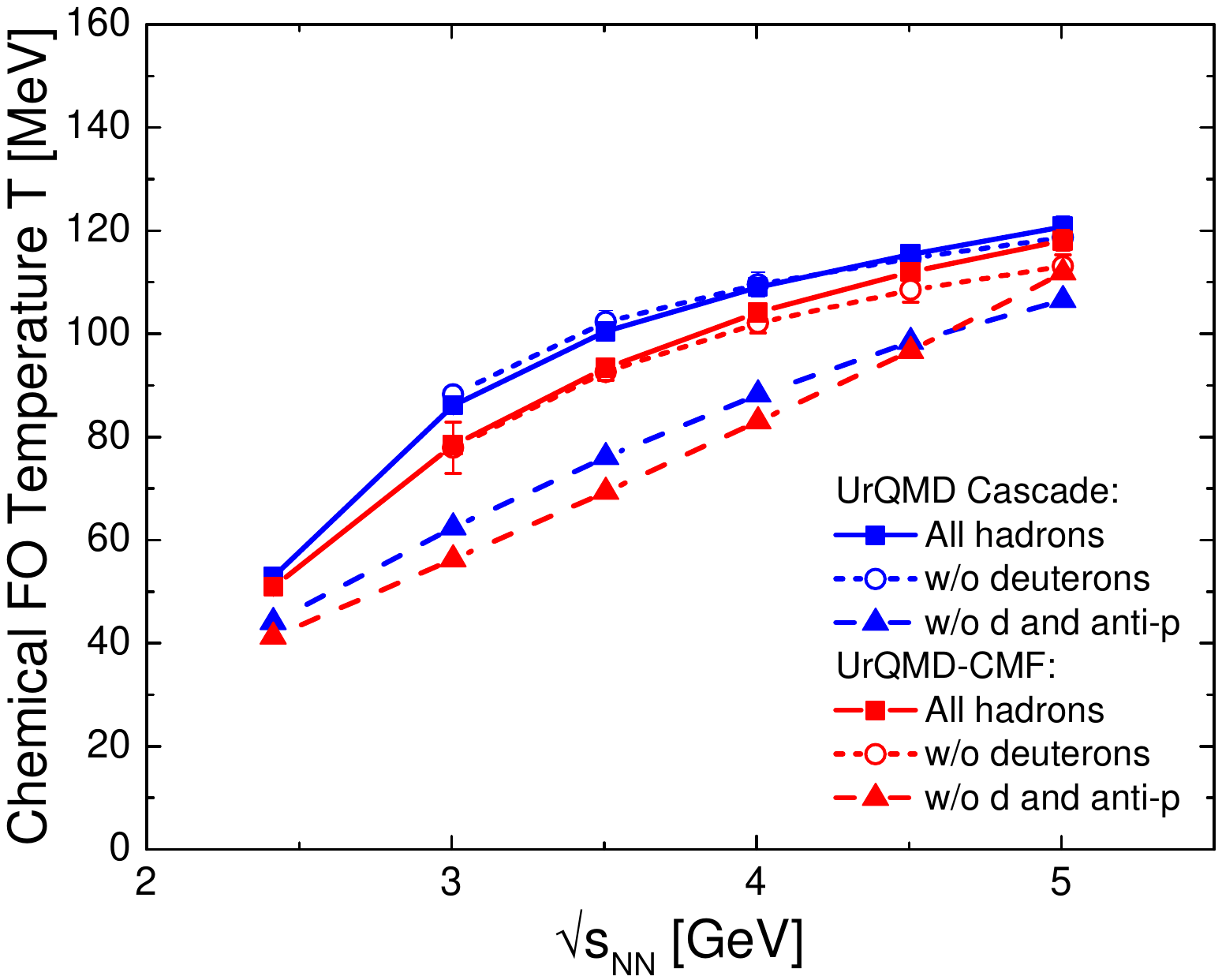}
    \includegraphics[width=0.49\textwidth]{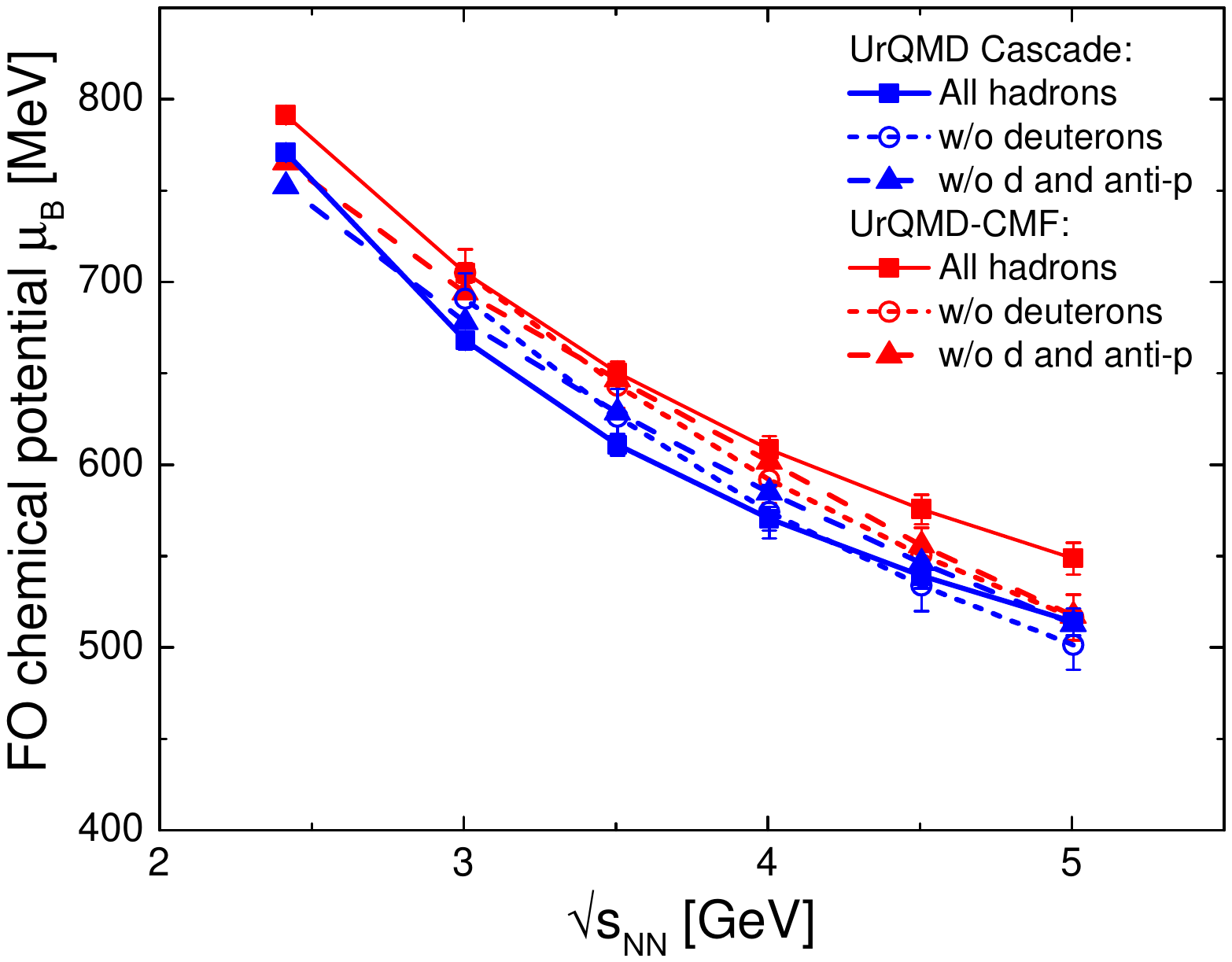}
    \caption{Thermal parameters (Temperature left and chemical potential right) from the FIST fits to the different UrQMD simulations as function of beam energy. Blue lines correspond to fits to the cascade mode and red lines to the CMF-potentials.  }
    \label{fig:thermal}
\end{figure*}

Finally, figures \ref{fig:thermal} summarize the differences in thermal parameters between the two equations of state and three different hadron sets directly as a function of the center of mass energy $\sqrt{s_{NN}}$. The line styles and colors are the same as in the two previous figures.

\section{Summary}

We have studied the dependence of thermal fit parameters on the choice of hadrons used as input for the fit as well as the underlying equation of state. We have found systematic effects from both which need to be taken into account when one wants to relate observables at certain beam energies to features of the QCD phase diagram using the thermal freeze-out line as reference.
Changes in the underlying equation of state may lead to a shift of $\approx 10$ MeV in the extracted temperature and $\approx 50$ MeV in the chemical potential. Even with an inconsistency in EoS between the thermal fit and underlying simulation (or measured data), a satisfactory fit can be obtained, albeit with systematically shifted freeze-out parameters. This is not a trivial result as it is not self-evident that such a good fit should still be possible with a change in the equation of state or that this would lead to the same freeze-out curve.
It also means that there is no unique (EoS-independent) relation between a beam energy and the expected freeze-out points.
The effects from an incomplete set of hadrons in the fit may even be larger. In addition, if a different equation of state is used for the fitting as for the evolution of the system, a shift in the beam energy at which a certain chemical freeze-out point is reached can be shifted by approximately 0.5 GeV which corresponds to more than $10\%$ of the beam energy in the center of mass frame. These systematic uncertainties need to be taken seriously if one wants to locate important features of the phase diagram from heavy ion collisions.

\section*{Acknowledgments}
T.R. gratefully acknowledges financial support by the Fulbright U.S. Scholar Program, which is sponsored by the U.S. Department of State and the German-American Fulbright Commission. This article’s contents are solely the responsibility of the authors and do not necessarily represent the official views of the Fulbright Program, the Government of the United States, or the German-American Fulbright Commission. T.R. gratefully acknowledges support from The Branco Weiss Fellowship - Society in Science, administered by the ETH Z\"urich. 
V.V. has been supported by the U.S. Department of Energy, 
Office of Science, Office of Nuclear Physics, Early Career Research Program under Award Number DE-SC0026065.
The computational resources for this project were provided by the Center for Scientific Computing of the GU Frankfurt and the Goethe--HLR and GSI green cube. The publication is funded by the Open Access Publishing Fund of GSI Helmholtzzentrum fuer Schwerionenforschung.

\bibliography{sn-bibliography}

\end{document}